\documentclass[conference]{IEEEtran}
\IEEEoverridecommandlockouts
\usepackage{cite}
\usepackage{amsmath,amssymb,amsfonts}
\usepackage{algorithmic}
\usepackage{graphicx}
\usepackage{textcomp}
\usepackage{xcolor}
\usepackage{float} 
\usepackage{subfig} 
\usepackage{url} 
\usepackage{hyperref}
\hypersetup{colorlinks=true,citecolor=black,filecolor=black,linkcolor=black,urlcolor=black,breaklinks=true}

\begin{document}

\makeatletter
    \newcommand{\linebreakand}{%
      \end{@IEEEauthorhalign}
      \hfill\mbox{}\par
      \mbox{}\hfill\begin{@IEEEauthorhalign}
    }
\makeatother

\title{Using Image Processing Techniques to Identify and Quantify Spatiotemporal Carbon Cycle Extremes\\

\thanks{This research was supported by the Reducing Uncertainties in Biogeochemical Interactions through 
Synthesis and Computation (RUBISCO) Science Focus Area, which is sponsored by the Regional and 
Global Model Analysis (RGMA) activity of the Earth \& Environmental Systems Modeling (EESM) Program 
in the Earth and Environmental Systems Sciences Division (EESSD) of the Office of Biological and 
Environmental Research (BER) in the US Department of Energy Office of Science.
This research used resources of the National Energy Research Scientific Computing Center (NERSC), 
a U.S. Department of Energy Office of Science User Facility located at Lawrence Berkeley National 
Laboratory, operated under Contract No. DE-AC02-05CH11231 for the Project m2467.
\\
}
}

\newcommand\blfootnote[1]{%
	\begingroup
	\renewcommand\thefootnote{}\footnote{#1}%
	\addtocounter{footnote}{-1}%
	\endgroup
}

\author{\IEEEauthorblockN{1\textsuperscript{st} Bharat Sharma}
\IEEEauthorblockA{\textit{Computational Earth Sciences Group} \\
\textit{Oak Ridge National Laboratory}\\
Oak Ridge, USA \\
bharat.sharma.neu@gmail.com}
\\
\IEEEauthorblockA{\textit{Department of Civil and Environmental Engineering} \\
\textit{Northeastern University}\\
Boston, USA \\
orcid.org/0000-0002-6698-2487}
\and
\IEEEauthorblockN{2\textsuperscript{nd} Jitendra Kumar}
\IEEEauthorblockA{\textit{Environmental Sciences Division} \\
\textit{Oak Ridge National Laboratory}\\
Oak Ridge, USA \\
orcid.org/0000-0002-0159-0546}
\linebreakand 
\IEEEauthorblockN{3\textsuperscript{rd} Auroop R. Ganguly}
\IEEEauthorblockA{\textit{Department of Civil and Environmental Engineering} \\
\textit{Northeastern University}\\
Boston, USA \\
orcid.org/0000-0002-4292-4856}
\and
\IEEEauthorblockN{4\textsuperscript{th} Forrest M. Hoffman}
\IEEEauthorblockA{\textit{Computational Earth Sciences Group} \\
\textit{Oak Ridge National Laboratory}\\
Oak Ridge, USA \\
orcid.org/0000-0001-5802-4134}
}

\maketitle

\begin{abstract}
    Rising atmospheric carbon dioxide due to human activities through fossil fuel 
    emissions and land use changes have increased climate extremes such as heat 
    waves and droughts that have led to and are expected to increase the occurrence 
    of carbon cycle extremes. Carbon cycle extremes represent large anomalies in the 
    carbon cycle that are associated with gains or losses in carbon uptake. 
    Carbon cycle extremes could be continuous in space and time and cross political 
    boundaries. Here, we present a methodology to identify large spatiotemporal 
    extremes (STEs) in the terrestrial carbon cycle using image processing tools for feature detection.
    We characterized the STE events based on neighborhood structures that are 
    three-dimensional adjacency matrices for the detection of spatiotemporal 
    manifolds of carbon cycle extremes. We found that the area affected and 
    carbon loss during negative carbon cycle extremes were consistent with 
    continuous neighborhood structures. In the gross primary production data we used, 100 carbon cycle STEs accounted for 
    more than 75\% of all the negative carbon cycle extremes. This paper 
    presents a comparative analysis of the magnitude of carbon cycle STEs 
    and attribution of those STEs to climate drivers as a function of neighborhood structures 
    for two observational datasets and an Earth system model simulation.
\end{abstract}


\blfootnote{This manuscript has been authored by UT-Battelle, LLC, under contract DE-AC05-00OR22725 with the US Department of Energy (DOE). The US government retains and the publisher, by accepting the article for publication, acknowledges that the US government retains a nonexclusive, paid-up, irrevocable, worldwide license to publish or reproduce the published form of this manuscript, or allow others to do so, for US government purposes. DOE will provide public access to these results of federally sponsored research in accordance with the DOE Public Access Plan (http://energy.gov/downloads/doe-public-access-plan).}

\section{Introduction}

Increased production and use of fossil fuels and deforestation have led to 
an increase in the atmospheric concentration of greenhouse gases (GHGs),
most importantly
carbon dioxide (CO$_{2}$), methane, and nitrous oxide. 
The increased concentration of GHGs is driving a rise in the 
surface temperature of the Earth, amplifying climate variability, and 
increasing the occurrence of climate extremes \cite{Reichstein_2013_climate_ext}.
Terrestrial ecosystems have historically taken up about 30\% of 
anthropogenic CO$_{2}$ emissions via carbon accumulation in plant biomass and 
soils \cite{Friedlingstein_2022_GCB}.
The increased carbon fertilization and water use efficiency, and the lengthening 
of growing seasons are increasing terrestrial carbon uptake 
and limiting the rise in 
atmospheric concentration of CO$_{2}$ \cite{Schimel_2015_CO2}. 
However, exacerbating climate extremes over time---such as droughts, heat waves,
and fires---have the potential to reduce terrestrial carbon 
uptake \cite{Reichstein_2013_climate_ext, Frank_ClimateExtremes_CC_2015,
Flach_Climate_extreme_GPP_2020, Piao_2013_CC}.

Recent studies have focused on detecting and quantifying extremes 
in the impacted systems, such as carbon cycle extremes 
\cite{Zscheischler_GRL_2014,Sharma_2022_CarbonExtremesLULCC,Sharma_2022_NBPExtremes},
and attribution of those extremes to climate drivers. 
Recent studies have analyzed spatiotemporal continuous extremes in the carbon cycle 
using 3$\times$3$\times$3 voxels and found that a few of these continuous extremes 
represent most of the interannual variability in the terrestrial carbon cycle \cite{Zscheischler_2013_STE, Zscheischler_2014_IAV_GPP}.
However, whether extreme events in the carbon cycle are discrete or continuous is 
an open question.
While the answer to this question is beyond the scope of this paper, we analyzed the 
proximity of carbon cycle extremes to each other, in the spatiotemporal manifold of connected
grid cells, depending on the shape of a three dimensional
search cube or voxel.
This is the first study that has quantified and compared the characteristics of continuous and 
non-continuous extremes in the global carbon cycle.

We defined six unit neighborhood structures composed of isolated or 
non-continuous extremes that are (1) continuous only in time, (2) continuous only in space,
or (3) continuous in both space and time (Figure~\ref{f:Manifold_Building_Block}). 
We analyzed and compared the characteristics of non-continuous, constrained continuous, 
and continuous extremes in GPP for observational datasets and Earth system model (ESM) outputs.
We then calculated the representative climatic conditions for such carbon 
cycle extremes to attribute them to climate drivers. 
We also investigated the characteristics of the connected network of carbon 
cycle extremes, which indicates how close, in space and time, extremes in terrestrial ecosystems 
occur.
The objectives of this study were to a) use image processing techniques 
to detect spatiotemporal manifolds of extreme events in GPP,
referred to in this paper as spatiotemporal extremes (STEs), 
b) compare the characteristics of STEs identified in ESM 
simulation output with those from gridded observational datasets,
and c) attribute the STEs in GPP to climate drivers.

\vspace{1cm}
\section{Data}
\label{sec:Data}

We used two observation-based, up-scaled GPP data products, FluxANN and GOSIF, 
and Earth system model output from the Community Earth System Model version 2 (CESM2) from an historical simulation conducted for the sixth phase of the Coupled Model Intercomparison Project (CMIP6).
FLUXCOM provides global gridded carbon fluxes from two experimental setups, 
one with only remote sensing (``RS'') input drivers and the other with RS and 
meteorological drivers (``RS+METEO''). 
The FluxANN dataset is ``FLUXCOM RS+METEO'', which used CRUNCEPv6 climate reanalysis and an
Artificial Neural Network (ANN) \cite{Jung_2020_Fluxcom}.
FluxANN was produced at $0.5^{\circ} \times 0.5{^\circ}$ spatial and 
monthly temporal resolution, and it is available at \href{https://www.fluxcom.org/}{https://www.fluxcom.org/}.
The GOSIF dataset was produced using global Solar-induced chlorophyll fluorescence (SIF) 
from discrete Orbiting Carbon Observatory-2 (OCO-2) SIF soundings, 
remote sensing data from MODIS,
and meteorological reanalysis data \cite{Xiao_GOSIF_2019}. 
GOSIF was produced at $0.05^{\circ} \times 0.05{^\circ}$ spatial and 
8-day temporal resolution, and it is accessible at
\href{https://globalecology.unh.edu/data/GOSIF-GPP.html}{https://globalecology.unh.edu/data/GOSIF-GPP.html}.
For attribution of GPP STEs from observations, we used climate driver data from the fifth generation European Centre for Medium-Range Weather Forecasts (ECMWF) reanalysis (ERA5) product.
ERA5 is produced at $0.1^{\circ} \times 0.1{^\circ}$ spatial and hourly temporal resolution, and it is
accessible at \href{https://www.ecmwf.int/en/forecasts/datasets/reanalysis-datasets/era5}{https://www.ecmwf.int/en/forecasts/datasets/reanalysis-datasets/era5}.
 
The CESM2 simulation was performed at $0.9375^{\circ} \times 1.25{^\circ}$ spatial resolution, and
monthly output was used here.
CESM2 is a fully coupled global Earth system model composed of 
atmosphere, ocean, land, sea ice, and land ice components. 
The CESM2 simulation output can be downloaded from 
\href{https://www.cesm.ucar.edu/models/cesm2/}{https://www.cesm.ucar.edu/models/cesm2/}.

\vspace{1cm}
\section{Methods}

\vspace{.5cm}
\subsection{\textbf{Data Preprocessing}}

The spatial resolutions of observation based GPP-datasets are $0.5^{\circ}$ and $0.05^{\circ}$ for
FluxANN and GOSIF, respectively, and $0.1^{\circ}$ for the ERA5 climate data. 
For consistent comparison of extremes and attribution, GOSIF and ERA5 were conservatively 
regridded to $0.5^{\circ}$ spatial resolution
using TempestRemap \cite{Paul_TempestRemap_2015,Paul_TempestRemap_2016}.
Both observational datasets were aggregated to monthly average timeseries.
A common time period from 2001-01-01 to 2013-12-31 (156 months) was chosen and all datasets were
trimmed to this study period.

We used the GPP datasets from up-scaled observational data products (FluxANN and GOSIF)
and from CESM2 to quantify 
spatiotemporal extreme (STE) events in GPP, also referred to as carbon cycle STEs, in this paper.
These carbon cycle STEs represented anomalous gains or losses in GPP or total photosynthetic 
uptake. 
Recent studies \cite{Sharma_2022_CarbonExtremesLULCC,Sharma_2022_NBPExtremes} have found that 
the magnitude and frequency of negative carbon cycle extremes is expected to be larger than that of
positive extremes.
Hence, we focused our analysis on the detection and attribution of negative carbon cycle STEs, which 
potentially have larger impacts on the carbon cycle than positive extremes.
For attribution of negative STEs in the carbon cycle to climatic conditions, we chose surface air 
temperature (\emph{tas}) and precipitation (\emph{pr}) datasets.

Most variables in the Earth system have some possibly nonlinear trend and a modulated annual cycle.
We defined the nonlinear trend as signals in the time series that are comprised of 
return periods of 10 years and longer \cite{Sharma_2022_CarbonExtremesLULCC}.
The modulated annual cycle is composed of signals with return periods of 12 months and its harmonics.
To extract the nonlinear trend and modulated annual cycle from each variable, we used singular 
spectral analysis \cite{Golyandina_Book_20010123,Sharma_2022_CarbonExtremesLULCC}.

Data preprocessing is an important step that could have varying impact on results, especially
because the 
scope of spatiotemporal carbon cycle extremes in this study was global \cite{Zscheischler_2013_STE}.
We followed the data preprocessing steps as described by Zscheischler et al. \cite{Zscheischler_2013_STE}.
The anomalies in GPP were calculated by subtracting the trend and the annual cycle; thus, the reported
losses during negative carbon cycle extremes were absolute global losses in potential carbon uptake,
which allows for comparison 
across time.
For compatibility across space, the trend and annual cycle were subtracted
from \emph{tas}, and  
scaled by dividing by standard deviation at every grid cell.
For \emph{pr}, the trend was removed and the detrended \emph{pr} at every grid cell was 
normalized by dividing with total \emph{pr}.

\vspace{.5cm}
\subsection{\textbf{Detection of Carbon Cycle Extremes}}

\begin{figure}[htbp]
\centerline{\includegraphics[trim = {0.01cm 0.01cm 0.01cm .05cm},clip,width=0.95\columnwidth]{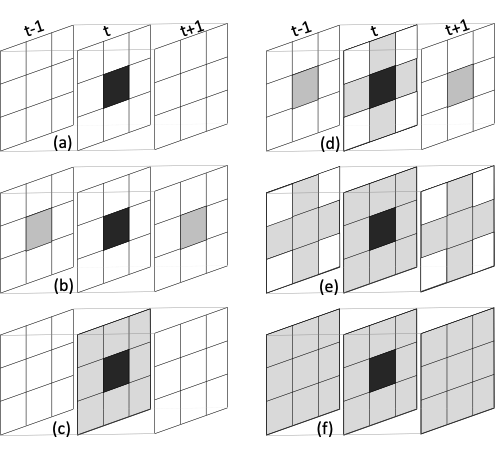}}
\caption{Different unit neighborhood structures of spatiotemporal extreme (STE) events. 
a)~sesd: small extent short duration; 
b)~seld: small extent long duration; 
c)~lesd: large extent short duration; 
d)~6-n: 6 neighbors;
e)~18-n: 18 neighbors; and 
f)~leld: large extent long duration or 26 neighbors}
\label{f:Manifold_Building_Block}
\end{figure}

Based on the global probability distribution of GPP anomalies, 
we selected the 10$^\mathrm{th}$ percentile value \cite{Seneviratne_IPCC_2012}, $q_{10}$,
such that total positive and negative extremes constitute 10\% of all GPP
anomalies \cite{Sharma_2022_NBPExtremes}.
GPP anomalies less than $q_{10}$ were called negative carbon cycle extremes,
and a three-dimensional (3D) mask of GPP extremes was calculated to produce data cubes of 1 and 0 values.
A value of 1 and 0 represented occurrence and non-occurrence of negative carbon cycle extremes, respectively.
We defined six neighborhood structures as shown in the Figure~\ref{f:Manifold_Building_Block},
that were used to search for the connected 3D manifold of carbon cycle extremes  \cite{Sharma_2022_IEEE}.
To find connected spatiotemporal extremes (STEs), we used the \emph{ndarray} library 
in Python. 
\emph{ndarray} connects the neighborhood structures using the adjacency matrix and assigns 
unique labels to each manifold.

\vspace{.5cm}
\subsection{\textbf{Attribution of Spatiotemporal Carbon Cycle Extremes}}

We quantified the response variable, GPP, within STE manifolds, then we 
computed the climatic conditions that could have
driven such STEs in GPP.
We calculated the representative climatic conditions using 
the median of the climate drivers throughout the spatiotemporal masks 
of the corresponding negative STEs in GPP.
Since terrestrial vegetation has an innate plastic capacity 
to buffer the impact of short duration climate extremes \cite{Zhang_Lag_effects_2014},
we computed the median of the climate drivers ($tas_{STE,t}$, $pr_{STE,t}$) during and 
up to three months ($N$)
prior to the occurrence of STEs in GPP.
We assumed that the \emph{tas} anomalies larger than the 75$^\mathrm{th}$ percentile represented
hot climatic conditions and \emph{pr} anomalies smaller than the 25$^\mathrm{th}$ percentile
represented dry conditions.
We performed the attribution analysis on the largest 100 STEs in GPP such that 
the number of negative carbon cycle STEs driven by $dry$, $wet$, $cold$, or $hot$ were
computed as shown in equations here,

\begin{equation}
    \#  dry = pr_{STE,t} < pr_{25q,t} \label{eq_dry}    ~~|~~ t \in N
\end{equation}

\begin{equation}
    \# wet = pr_{STE,t} > pr_{75q,t} \label{eq_wet}   ~~|~~ t \in N
\end{equation}

\begin{equation}
    \#  cold = tas_{STE,t} < tas_{25q,t} \label{eq_cold}    ~~|~~ t \in N
\end{equation}

\begin{equation}
    \# hot = tas_{STE,t} > tas_{75q,t} \label{eq_hot}   ~~|~~ t \in N
\end{equation}
 
\noindent
where $25q$ and $75q$ indicate the 25$^\mathrm{th}$ and 75$^\mathrm{th}$ percentile values 
of a climate driver at $t$ months prior to the occurrence of STE events.
$N$ is the total number of months (here, $N=3$) considered for lagged response of GPP on the climate drivers.

\vspace{.5cm}
\subsection{\textbf{Scale-Free Property of Carbon Cycle STEs}}

In a scale-free network, degree distribution of the nodes of the
network follows a power law \cite{Barabasi_2013_Network_Science},
such that the probability of a randomly chosen node, $n$, is 
$p(n)$ and has $\gamma$ links to other nodes.

\begin{equation}
    p(n) = C n^{-\gamma}   \label{eq_powerlaw_1}
\end{equation}

\begin{equation}
    p(n) = \log C + -{\gamma}\log n \label{eq_powerlaw_2}
\end{equation}

Many networks exhibit the scale-free property, such as, the Internet, actor networks, air traffic networks, etc. 
For example, in an air traffic network, most large airports act 
as hubs and have a large number of links to other large hubs and 
small airports \cite{Warner_2019_Network_Science}. 
Thus, the degree distribution in a scale-free network has a few 
nodes with large degree and most nodes with a small degree. 
This is in contrast to a random network of, say, state highways, 
in which we have a small number of highways for every city and, hence, 
the degree distribution per node is similar to the average
degree distribution of the entire network.
Similar scale-free properties have been observed in the disturbance of extreme events 
in terrestrial ecosystems as well \cite{Zscheischler_2013_STE}.
Since the STEs in the carbon cycle are a large scale connected component 
of extremes, we performed the power law fit for STEs in GPP for 
both FluxANN and CESM2 for multiple neighborhood structures.

\vspace{1cm}
\section{Results and Discussion}

\begin{figure}
 \subfloat{\includegraphics[trim = {0.01cm 0.01cm 0.01cm 0.25cm},
 clip, width=0.95\columnwidth]{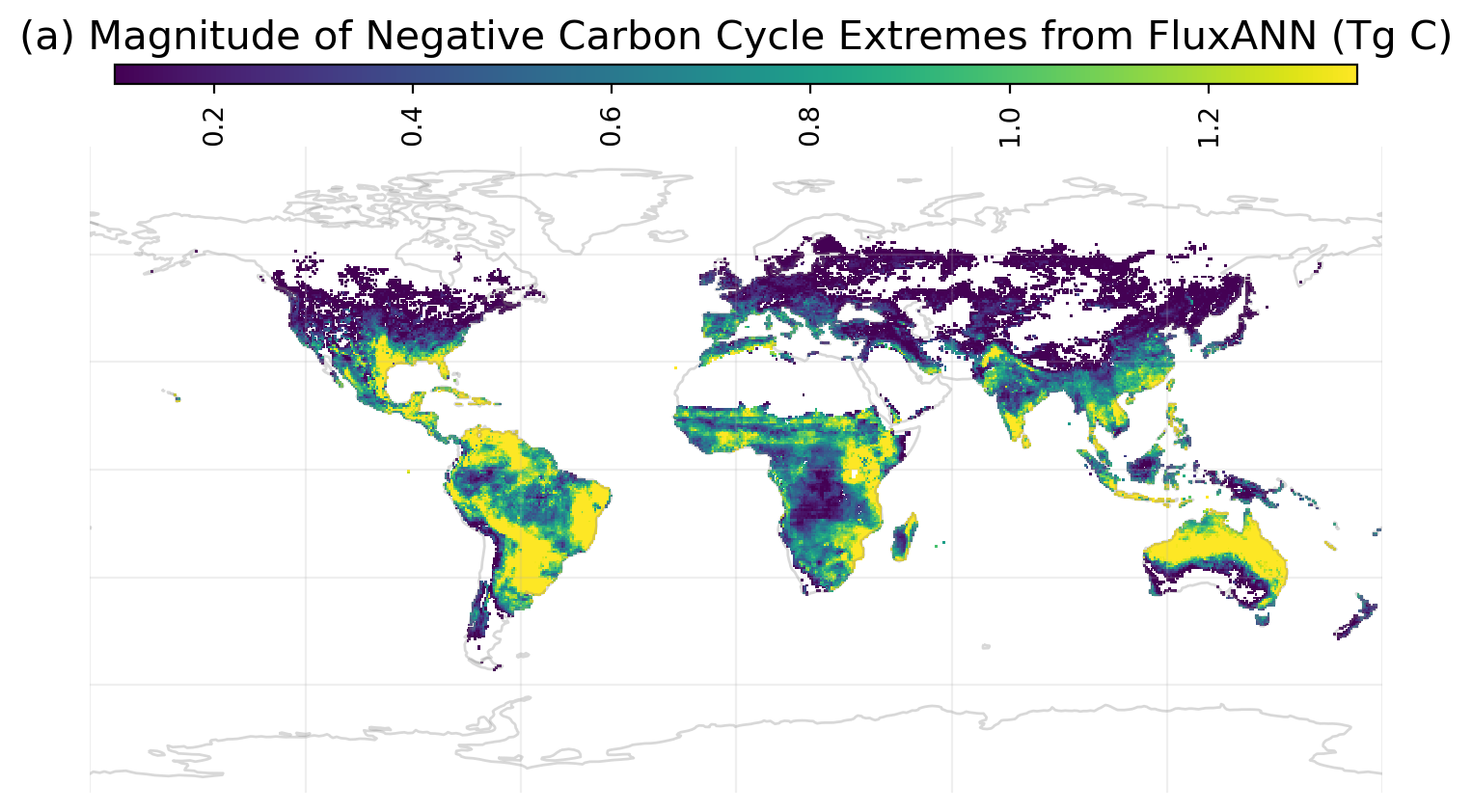}
 \label{f:spatial_fluxann}} \\
 \subfloat{\includegraphics[trim = {0.01cm 0.01cm 0.01cm 0.25cm},
 clip,width=0.95\columnwidth]{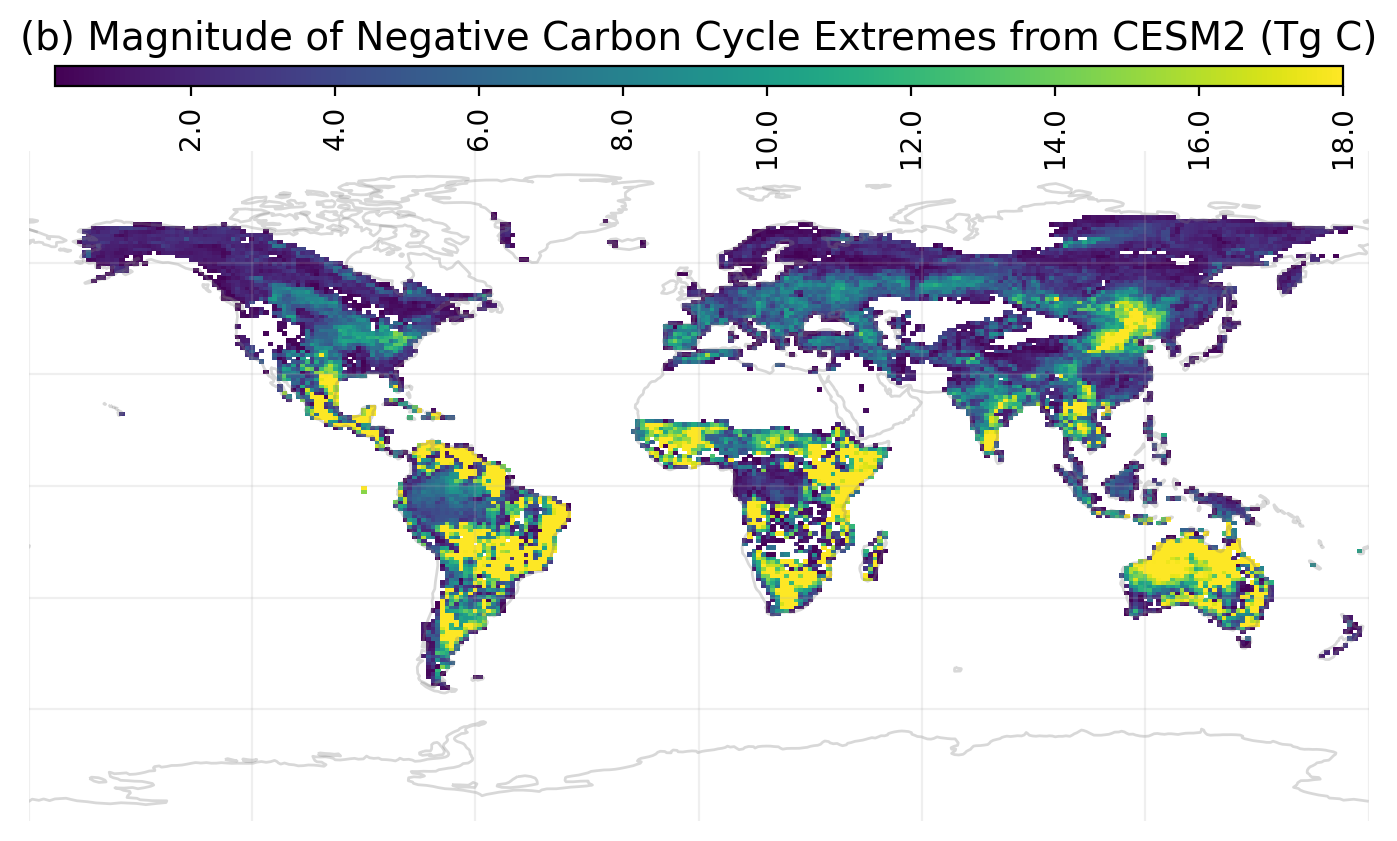}
 \label{f:spatial_cesm2}}\\

 \caption{The figure shows the spatial distribution of the integrated sum
 of carbon uptake loss (Tg\,C) during
 negative carbon
 cycle extremes for (a) FluxANN and (b) CESM2.}
 \label{f:spatial_neg_ext}
\end{figure}

\vspace{.5cm}
\subsection{\textbf{Characteristics of Carbon Cycle STEs }}
The annual mean magnitude of total GPP of terrestrial ecosystem 
was about 115~Pg\,C\,yr$^{-1}$ for both FluxANN and CESM2 from 
2001 to 2013.
However, the annual average loss in carbon uptake during negative
carbon cycle extremes was 1.6~Pg\,C\,yr$^{-1}$ for FluxANN and 
5.5~Pg\,C\,yr$^{-1}$ for CESM2. 
The larger magnitude of negative carbon cycle extremes is largely
driven by a larger amplitude of interannual variability (IAV) in the terrestrial carbon cycle 
\cite{Sharma_2022_CarbonExtremesLULCC}.
This implies that CESM2 was overestimating the IAV in GPP,
FluxANN was underestimating the IAV in GPP, or both.
Jung et al. \cite{Jung_2020_Fluxcom} investigated the 
bias and interannual variability metrics of FLUXCOM and concluded that 
FLUXCOM and FluxANN have a stronger carbon uptake strength in the tropics 
due to a positive bias in the underlying eddy covariance data.
Moreover, the IAV in GPP was weaker in FLUXCOM than in dynamic global 
vegetation models.
The main reason for the underestimation of IAV in FLUXCOM is due 
to the sparsity of observational sites, lack of 
site historical data, and smoothing by machine learning 
models \cite{Jung_2020_Fluxcom}.

\begin{figure}
	\begin{center}
 \subfloat{\includegraphics[trim = {0.01cm 0.01cm 0.01cm 0.35cm} , clip, width=0.80\columnwidth]{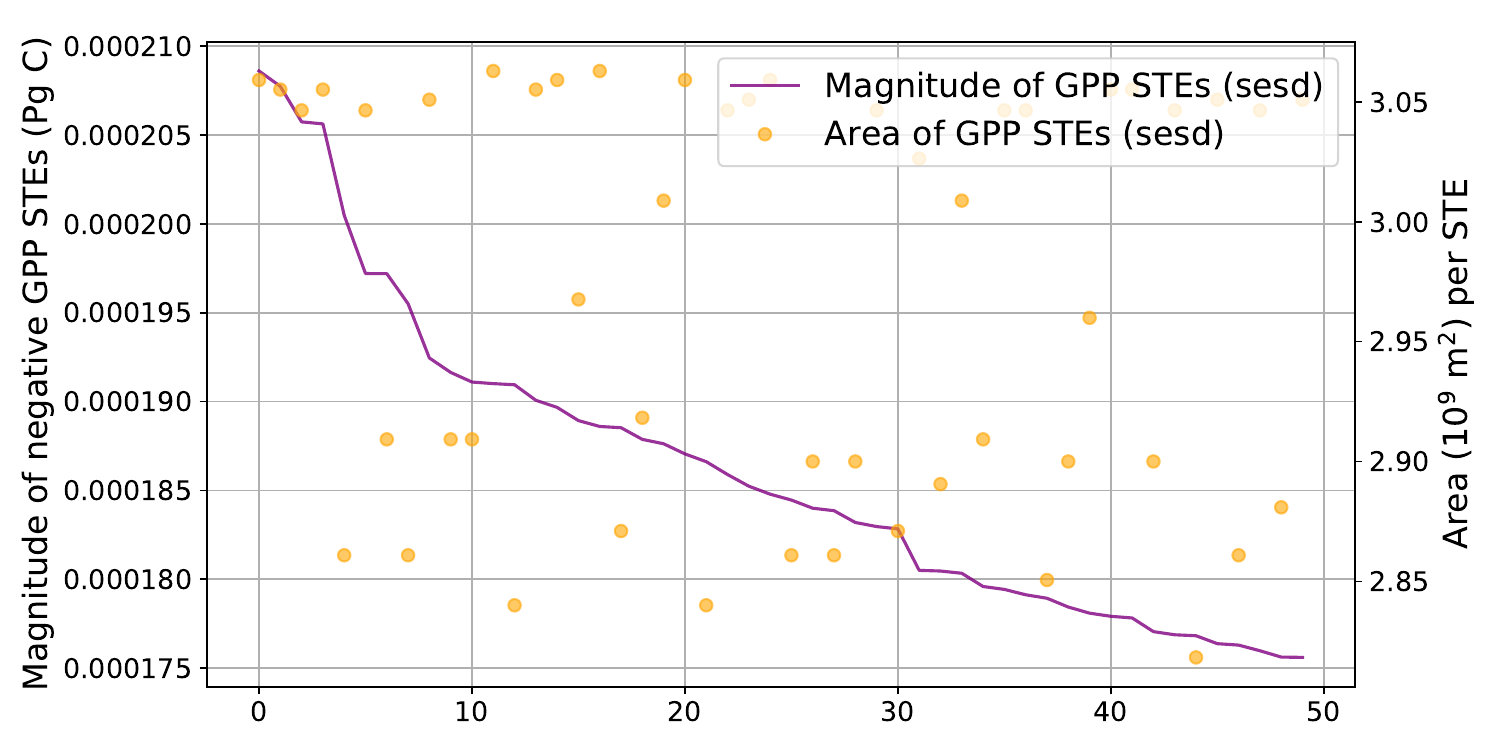}
 \label{f:FluxAnn_loss_area_sesd}} \\
 \vskip-0.093cm
  \subfloat{\includegraphics[trim = {0.01cm 0.01cm 0.01cm 0.35cm} , clip,width=0.80\columnwidth]{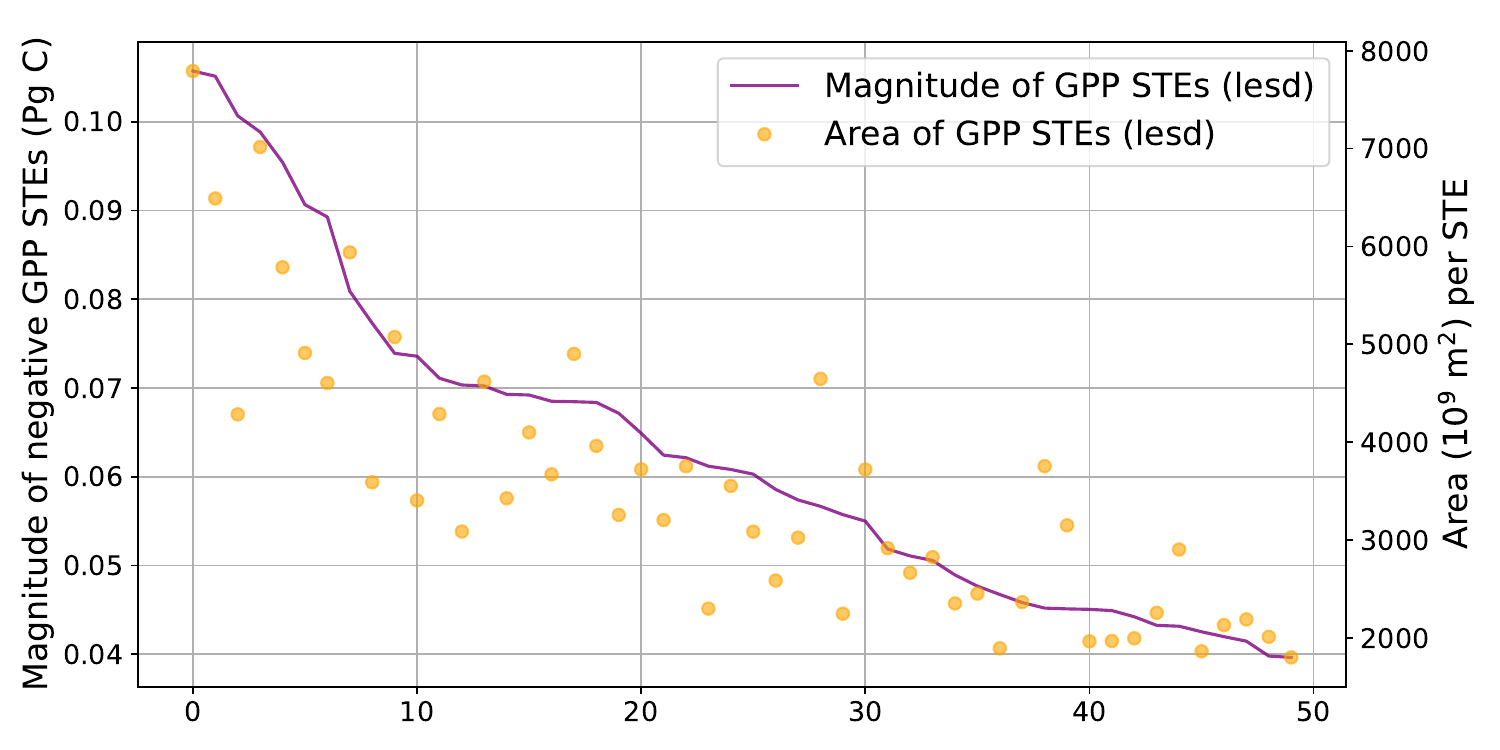}
 \label{f:FluxAnn_loss_area_lesd}} \\
 \vskip-0.093cm
 \subfloat{\includegraphics[trim = {0.01cm 0.01cm 0.01cm 0.35cm} , clip,width=0.80\columnwidth]{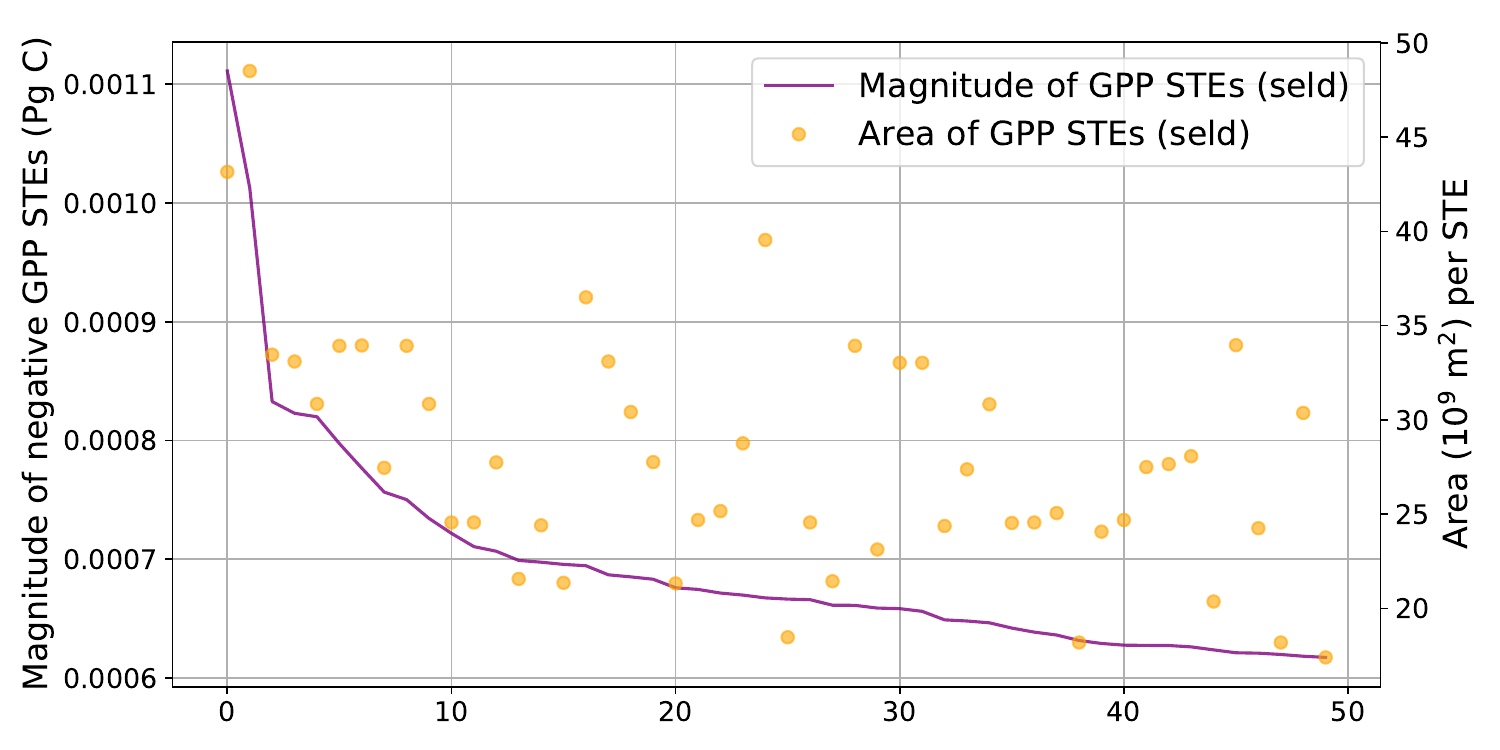}
 \label{f:FluxAnn_loss_area_seld}} \\
 \vskip-0.093cm
 \subfloat{\includegraphics[trim = {0.01cm 0.01cm 0.01cm 0.35cm} , clip, width=0.80\columnwidth]{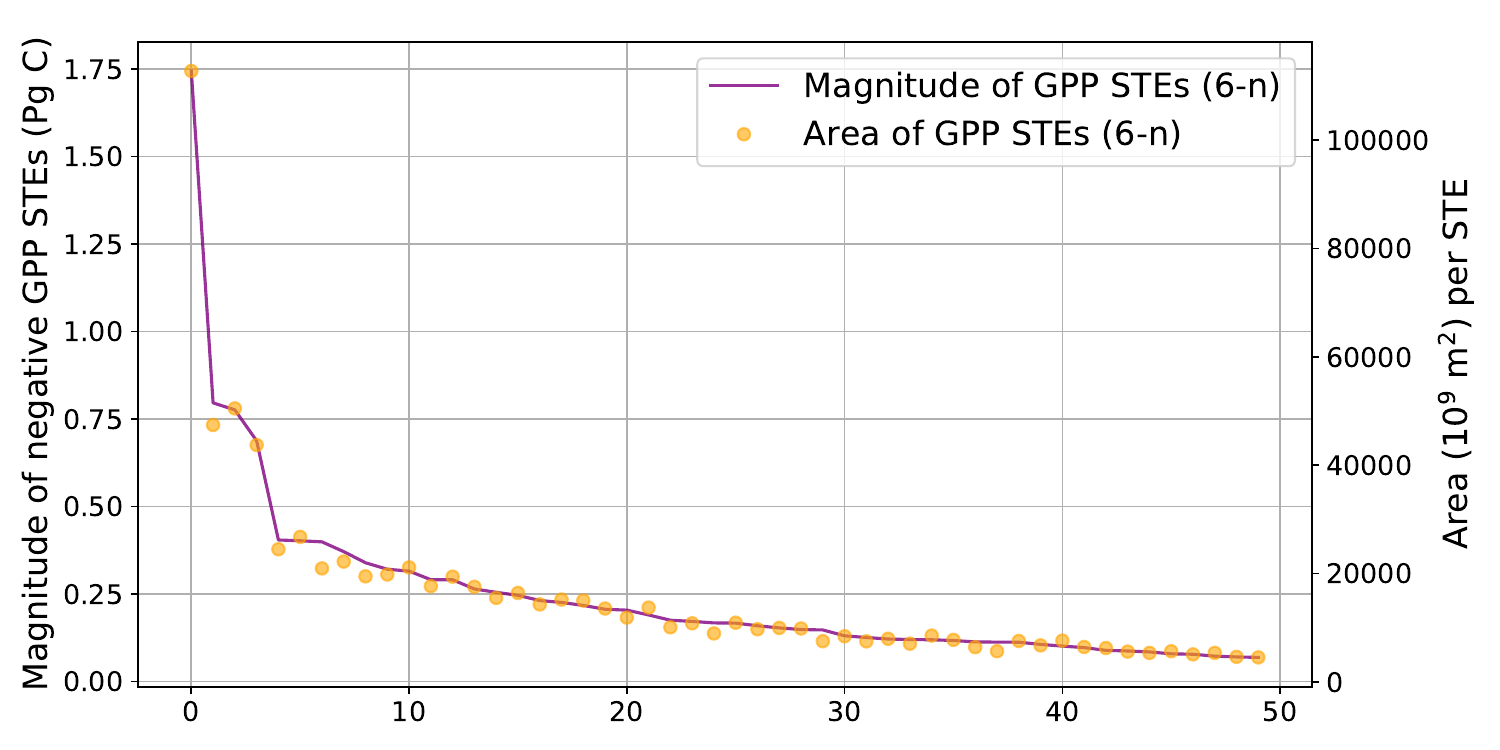}
 \label{f:FluxAnn_loss_area_6-n}} \\
 \vskip-0.093cm
  \subfloat{\includegraphics[trim = {0.01cm 0.01cm 0.01cm 0.35cm} , clip,width=0.80\columnwidth]{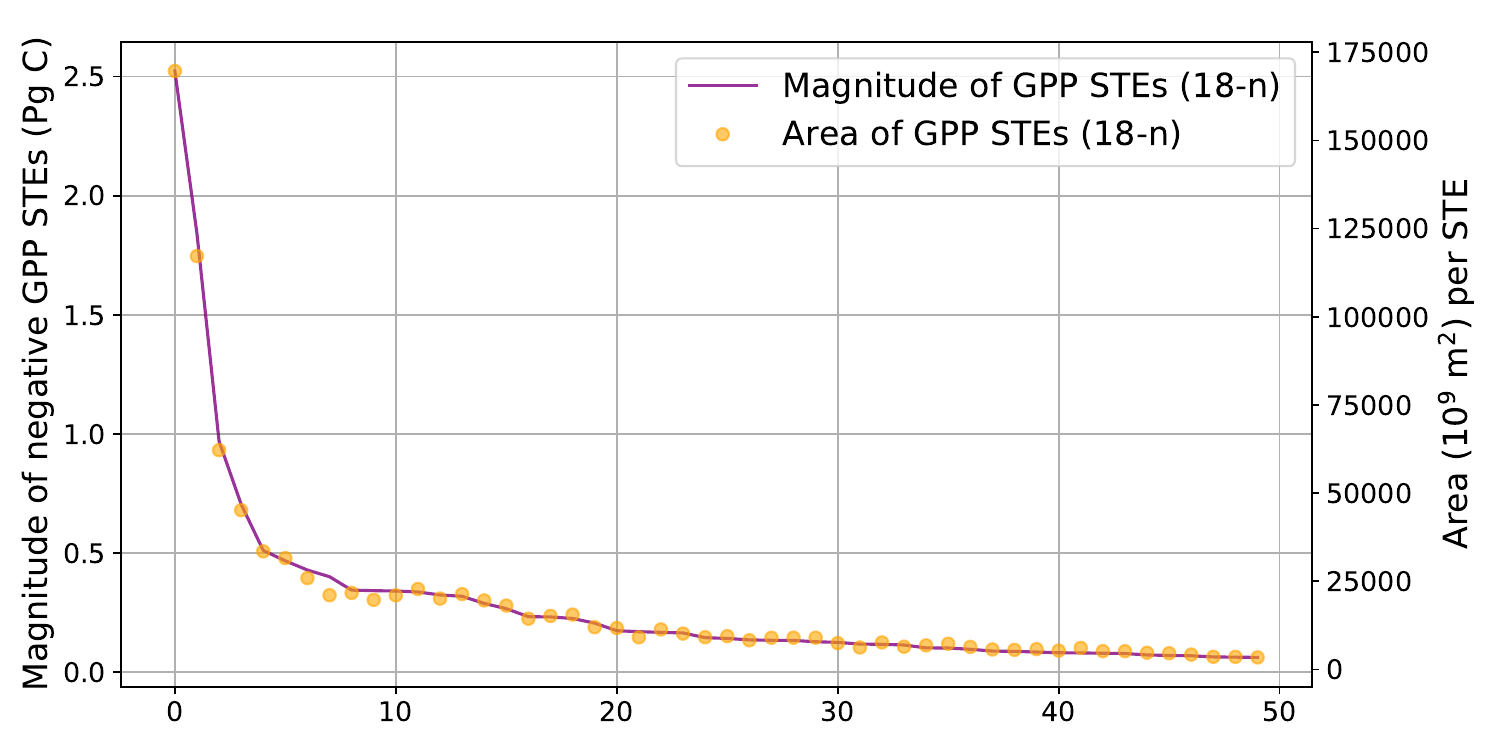}
 \label{f:FluxAnn_loss_area_18-n}} \\
 \vskip-0.093cm
 \subfloat{\includegraphics[trim = {0.01cm 0.01cm 0.01cm 0.35cm} , clip,width=0.80\columnwidth]{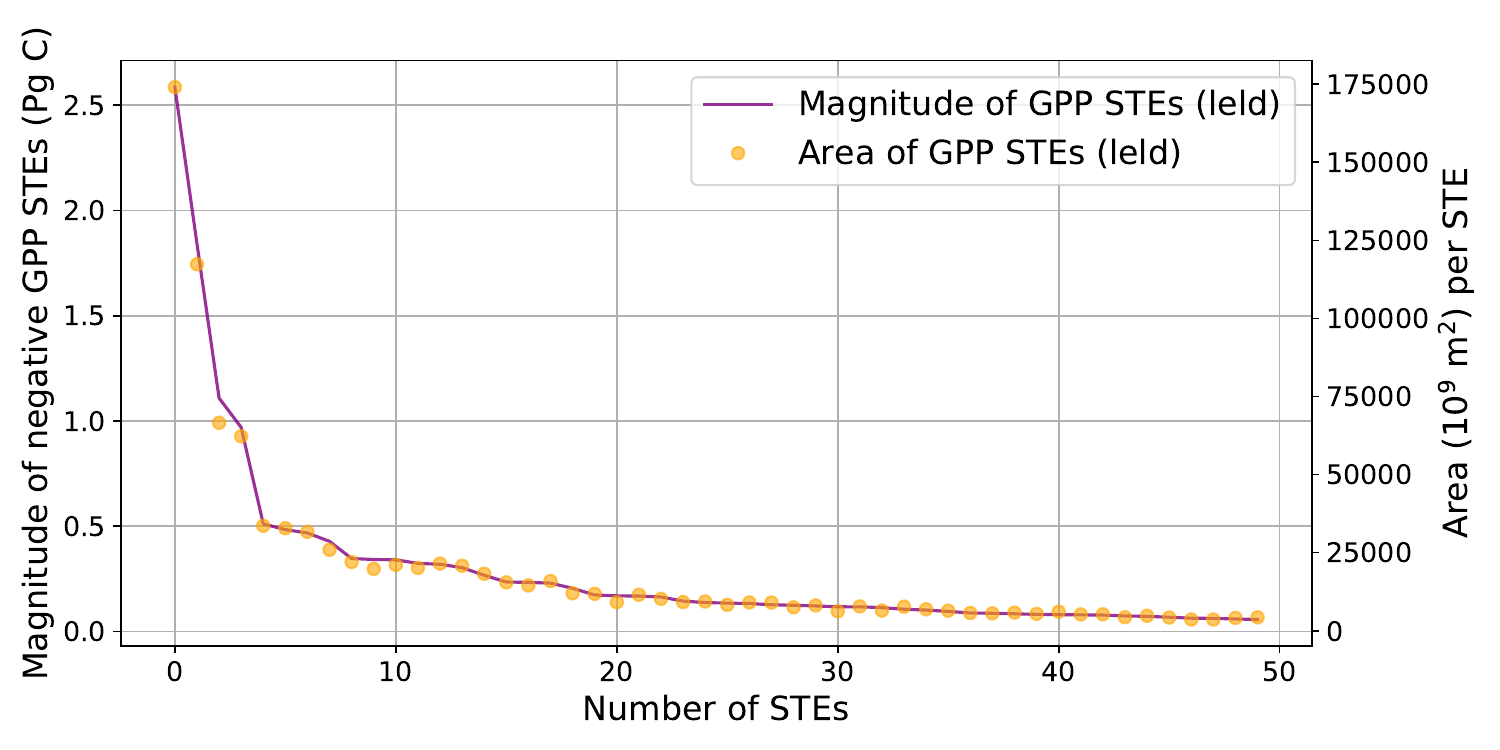}
 \label{f:FluxAnn_loss_area_leld}}
 \end{center}

 \caption{Plots of the loss of carbon uptake (Pg\,C) for each negative carbon cycle STE geometry
 for FluxANN, 
 shown as a line plot (left $y$-axis).
 The filled circles represent the corresponding affected area (10$^{+9}$~m$^{2}$) for each 
 negative carbon cycle STE geometry (right $y$-axis).
 The $x$-axis represents the top 50 negative carbon cycle STEs in 
 descending order of magnitude.}
 \label{f:gpp_loss_area_fluxann}
\end{figure}

The spatial distribution of 
the magnitude of negative carbon cycle extremes for FluxANN 
(Figure~\ref{f:spatial_fluxann}) and CESM2 (Figure~\ref{f:spatial_cesm2}) 
indicate large agreement on the spatial distribution of 
negative carbon cycle extremes, but disagreement on the magnitude of 
those extremes. 
The regions that show the largest magnitude of negative carbon cycle 
extremes were the Amazon Basin, Central and Southern South America, Eastern Africa, 
Eastern China, and Northern Australia. 
Due to low IAV in GPP in FluxANN \cite{Jung_2020_Fluxcom}, the magnitude of 
negative carbon cycle 
extremes was smaller compared to the negative carbon 
cycle extremes simulated by CESM2.

The area affected and magnitude of STEs in GPP increased with the size 
of the neighborhood structure. 
Figure~\ref{f:gpp_loss_area_fluxann} shows the distribution of the magnitude of 
the negative carbon cycle STEs in FluxANN in decreasing order of magnitude
and the corresponding area affected during STEs.
The largest magnitude of negative carbon cycle extremes with the \emph{sesd} structure was 
$2 \times 10^{-4}$~Pg\,C (Figure~\ref{f:gpp_loss_area_fluxann}a), 
whereas the largest negative STE with \emph{leld} structure was 2.5~Pg\,C 
(Figure~\ref{f:gpp_loss_area_fluxann}f).
With increasing size of neighborhood structure, the area and magnitude of 
STEs showed a consistent relationship with the distribution of ranked STEs.

\begin{figure}
	\begin{center}
 \subfloat[FluxANN]{\includegraphics[trim = {0.01cm 0.01cm 0.01cm 0.95cm} , clip, width=0.95\columnwidth]{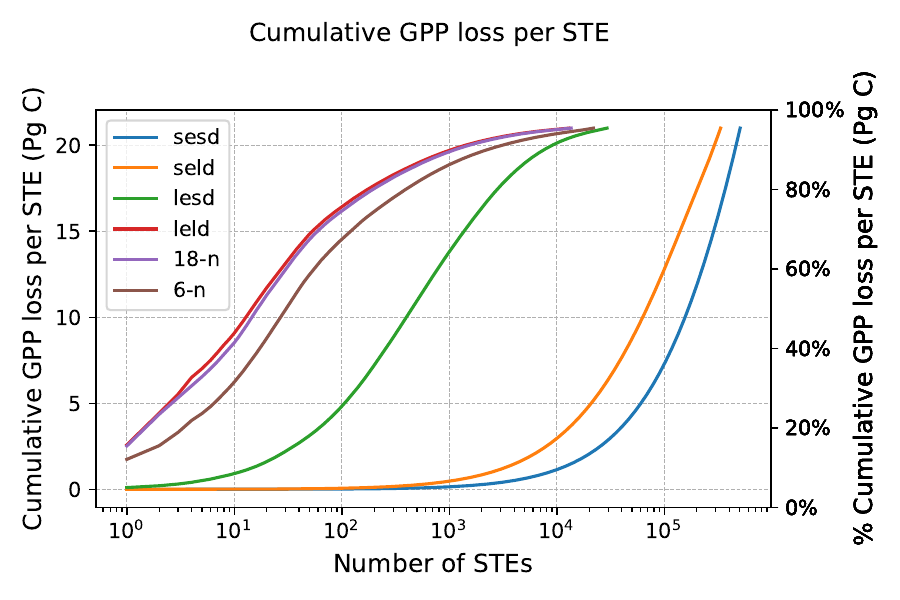}
 \label{f:gpp_cum_loss_stes_fluxann}} \\
 \subfloat[CESM2]{\includegraphics[trim = {0.01cm 0.01cm 0.01cm 0.95cm} , clip,width=0.95\columnwidth]{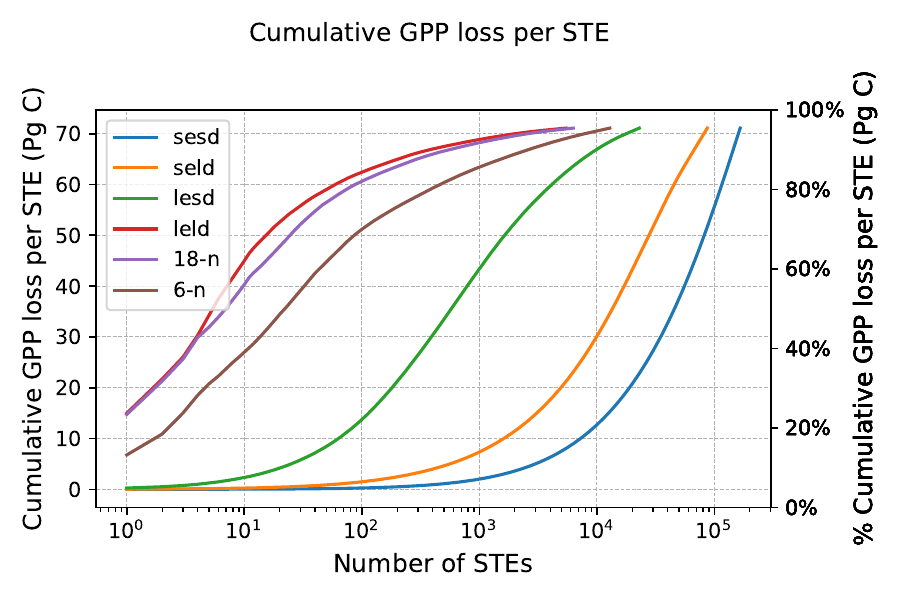}
 \label{f:gpp_cum_loss_stes_cesm2}}\\
  \subfloat[GOSIF]{\includegraphics[trim = {0.01cm 0.01cm 0.01cm 0.95cm} , clip,width=0.95\columnwidth]{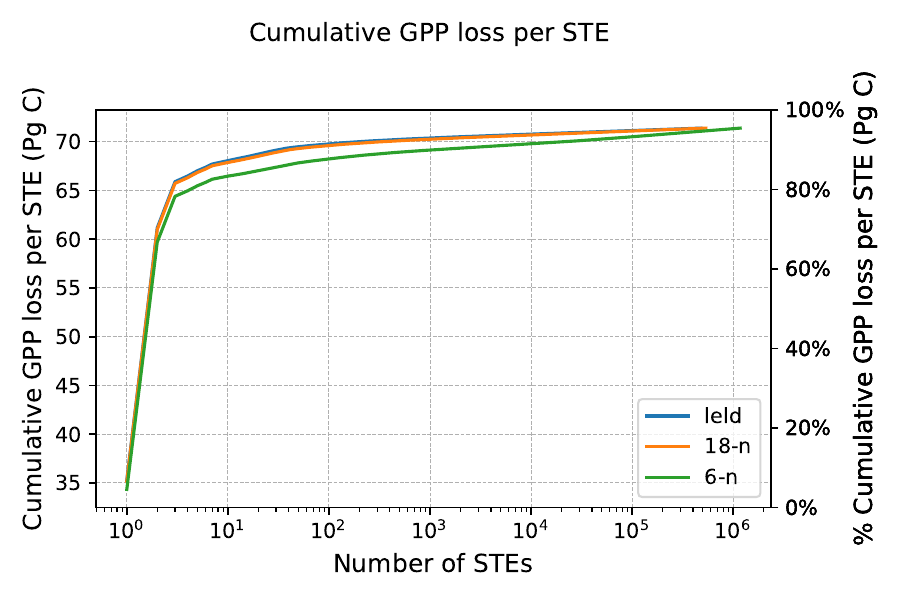}
 \label{f:gpp_cum_loss_stes_gosif}}\\
 \end{center}

 \caption{Plot of cumulative loss of carbon uptake during negative carbon cycle STEs.
 The distribution of magnitude (left $y$-axis) and percent carbon uptake 
 loss (right $y$-axis)
 are shown in 
 descending order of the magnitude of STEs.
 These distributions are shown for all neighborhood structures 
 for (a) FluxANN and (b) CESM2. 
 For (c) GOSIF, the distribution of GPP loss vs. number of STEs is 
 shown only for \textit{leld}, \textit{18-n}, 
 and \textit{6-n}.}
 \label{f:cum_gpp_loss_vs_stes}
\end{figure}

Figure~\ref{f:cum_gpp_loss_vs_stes} shows the cumulative 
losses in carbon uptake (GPP, Pg\,C) with increasing number of STEs.
The top 10 STEs for \emph{leld} account for about 45\%, 65\% and 85\% of 
the total loss of carbon uptake for FluxANN, CESM2, and GOSIF, respectively.
The top 100 STEs for \emph{leld} account for more than 75\% of the total carbon 
losses, which demonstrates that only a few large STEs can explain most of the
variability and losses in GPP. 
This finding is consistent with the results of \cite{Zscheischler_2014_IAV_GPP}.
As the neighborhood structure size is reduced, a larger number of STEs are required to 
account for similar losses in carbon cycle uptake (Figures
~\ref{f:cum_gpp_loss_vs_stes}a and ~\ref{f:cum_gpp_loss_vs_stes}b).

The total cumulative losses in the observation-based datasets, FluxANN and GOSIF, were 
about 20.9~Pg\,C and 73.4~Pg\,C, respectively. 
The larger magnitude of negative carbon cycle extremes in GOSIF were due to 
the larger magnitude and point density of IAV in GPP (Figure~\ref{f:IAV_GPP}) with 
respect to FluxANN.
These differences also highlighted large disagreements among different observation-based datasets.
The spatial resolution of GOSIF is 10 times finer than FluxANN and likely represents 
more variance in the carbon cycle; the ANN used to up-scale point observations to a coarser
gridded dataset was smoothed and represents lower variance in the carbon cycle 
\cite{Jung_2020_Fluxcom}.

\begin{figure}[htbp]
\centerline{\includegraphics[trim = {0.01cm 0.01cm 0.01cm .05cm},clip,width=0.95\columnwidth]{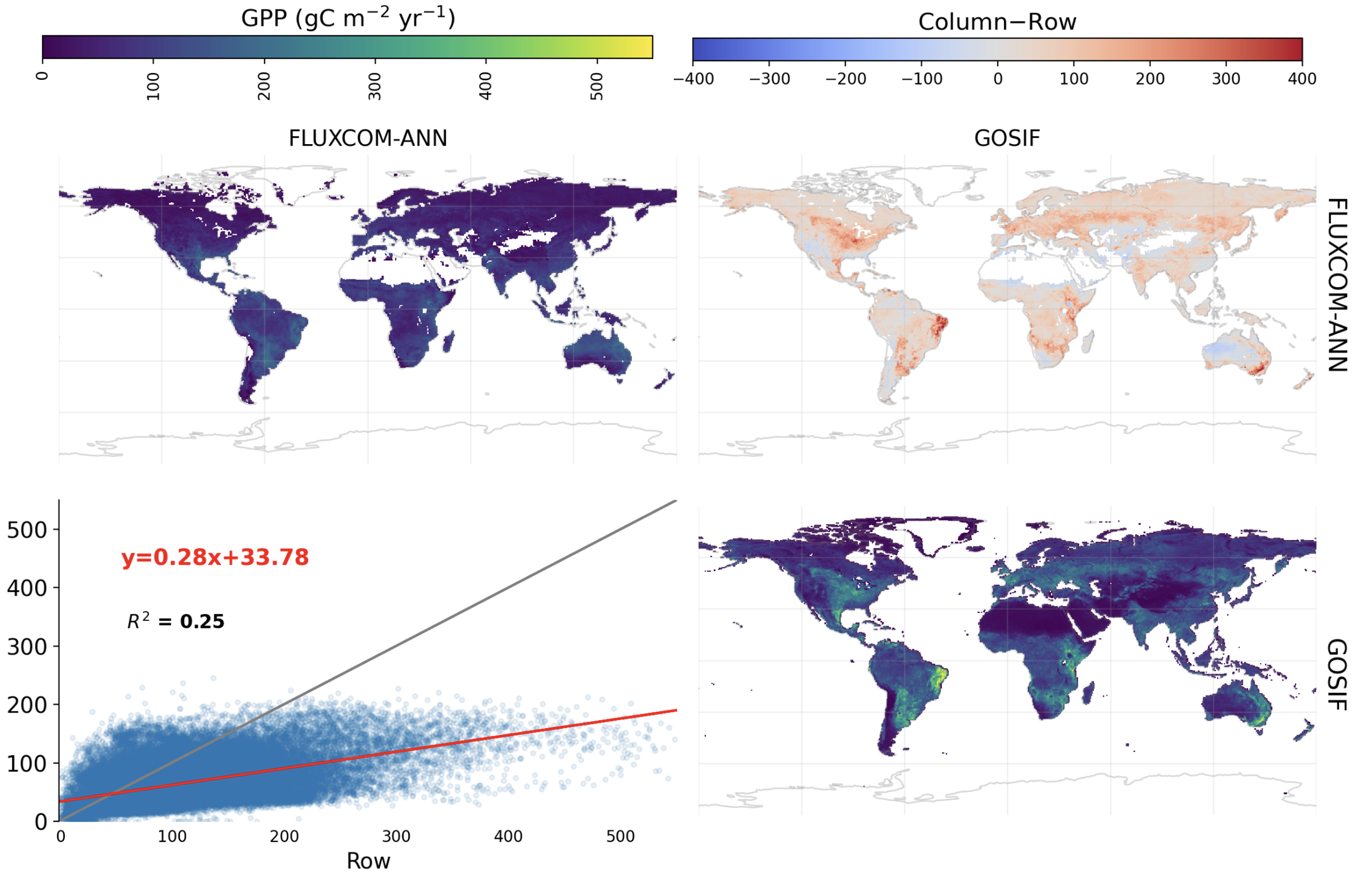}}
\caption{Left and right, and top and bottom, show the FLUXCOM-ANN (FluxANN) and GOSIF, respectively.
The diagonal maps show the IAV of GPP of both datasets.
The map above the diagonal shows the difference of the IAV of GPP of column dataset $-$ row dataset.
The map below the diagonal shows the point density in blue and 1:1 regression line in grey. 
Red line and equation represent the best fit line from total least-squares regression.}
\label{f:IAV_GPP}
\end{figure}

\vspace{.5cm}
\subsection{\textbf{Scale-free Property of STEs}}

Figure~\ref{f:powerlaw} shows the power law fit 
for the structure $leld$ for both FluxANN and CESM2. 
The exponent, $\gamma_{leld}$, for FluxANN and CESM2 was 1.83 
and 1.79, respectively, despite the different spatial resolutions 
of these datasets.
For every neighborhood structure that has continuity in both 
space and time, i.e., $6-n$, $18-n$, and $leld$, 
$1.75 < \gamma < 2$ as shown in Table~\ref{t:powerlaw},
which is also consistent with the literature 
\cite{Zscheischler_2014_IAV_GPP,Zscheischler_2013_STE}.

For a scale-free network, the natural cutoff is represented
as \cite{Barabasi_2013_Network_Science}

\begin{equation}
    n_{max} = n_{min} ~ M^{\frac{1}{\gamma - 1}} \label{eq_powerlaw_3}
\end{equation}

\noindent
where $M$ is the total number of nodes in the network and $n_{max}$,
$n_{min}$ are the degree of largest, smallest node, respectively.

\begin{figure}
 \subfloat[FluxANN]{\includegraphics[trim = {0.01cm 0.01cm 0.01cm 1.35cm} , clip, width=0.95\columnwidth]{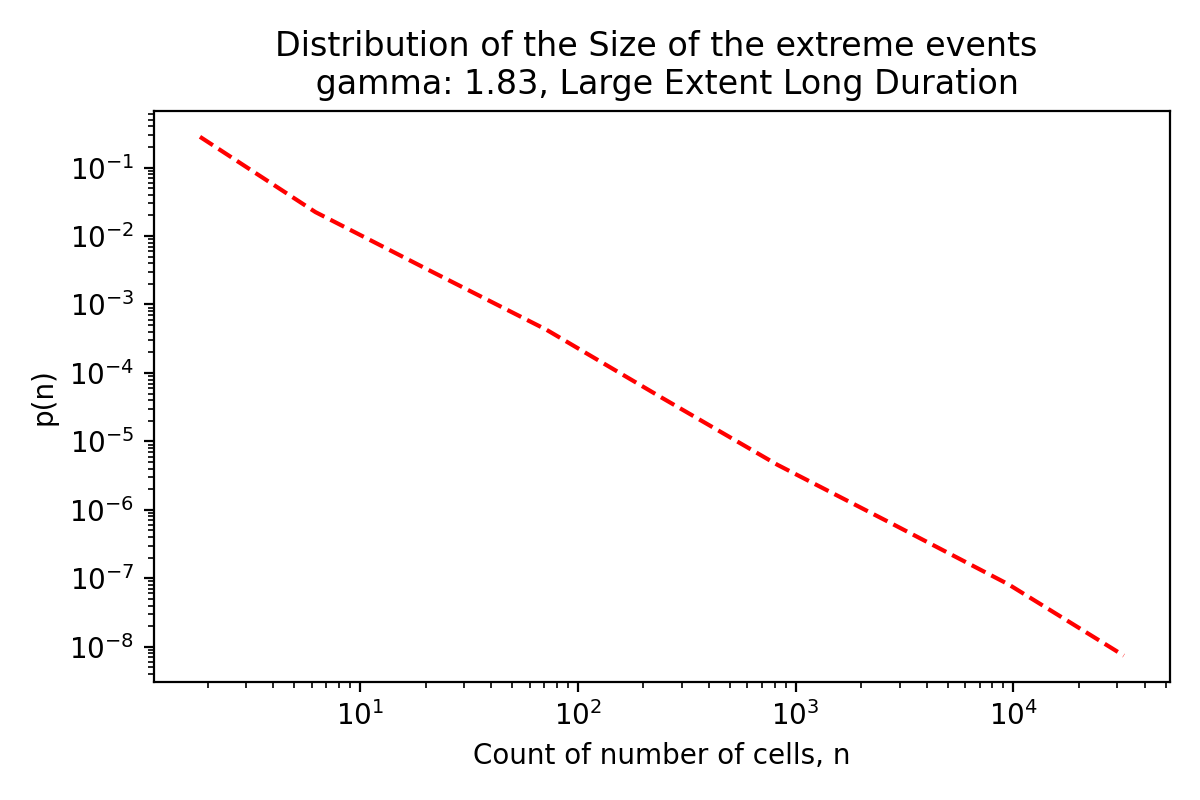}
 \label{f:powerlaw_cesm2_fluxann}} \\
 \subfloat[CESM2]{\includegraphics[trim = {0.01cm 0.01cm 0.01cm 1.35cm} , clip,width=0.95\columnwidth]{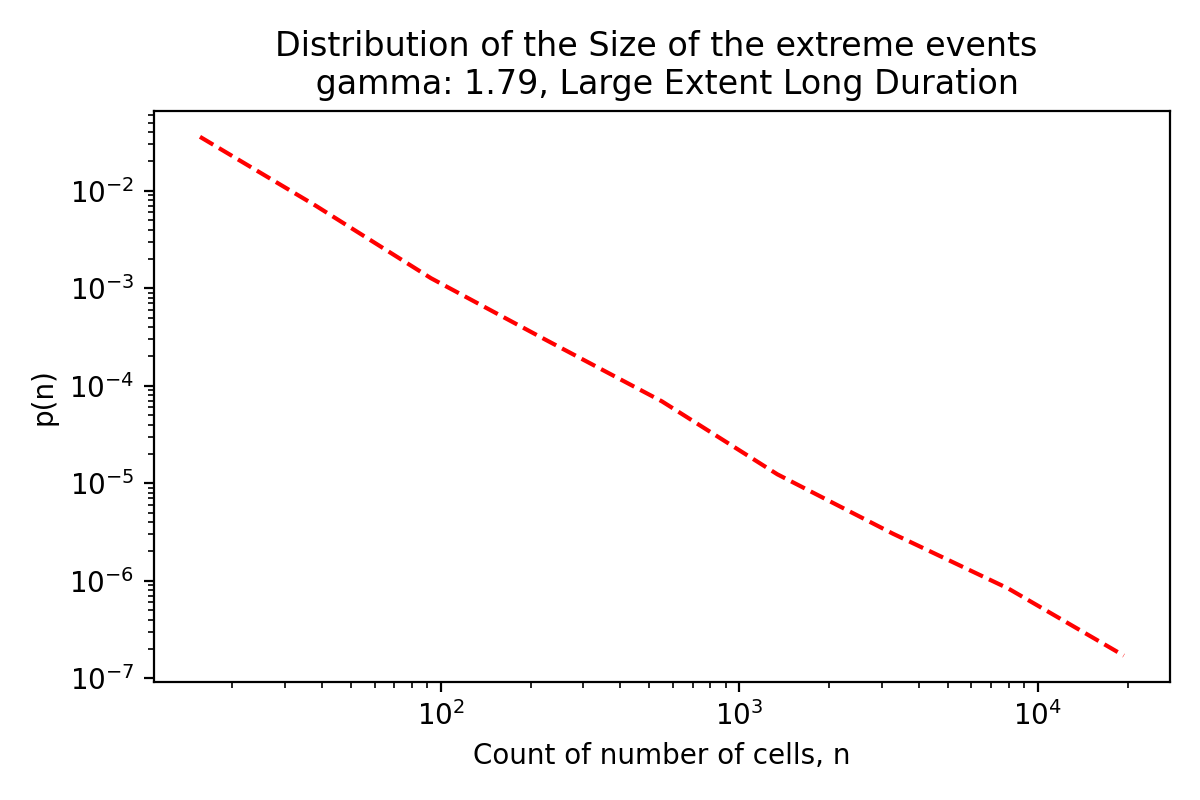}
 \label{f:powerlaw_cesm2_count}}\\

 \caption{Powerlaw fit for neighborhood structure 
 $leld$ for (a) FluxANN and (b) CESM2. 
 The $y$-axis represents the probability that randomly chosen
 manifold has $n$ degrees, see equation~\ref{eq_powerlaw_1}. 
 The $x$-axis represents the size of the manifold with $n$ nodes. 
 The corresponding power law exponent $\gamma$ for (a) FluxANN and (b) CESM2
 were 1.83 and 1.79, respectively.}
 \label{f:powerlaw}
\end{figure}

\begin{table}[]
    \centering
    \caption{Power law degree exponent ($\gamma$) for FluxANN and CESM2 for different neighborhood structures.}
    \label{t:powerlaw}
    \resizebox{.65\columnwidth}{!}{%
    \begin{tabular}{|c|r|r|}
    \hline
    \textbf{Exponent} & FluxANN & CESM2 \\ \hline
    \textit{seld}              & 20.68   & 6.77  \\ \hline
    \textit{lesd}              & 3.40     & 2.05  \\ \hline
    \textit{6-n}               & 1.85    & 1.88  \\ \hline
    \textit{18-n}              & 1.82    & 1.80   \\ \hline
    \textit{leld}              & 1.83    & 1.79  \\ \hline
    \end{tabular}%
    }
    \end{table}

For networks where $1 < \gamma < 2$, the exponent 1/($\gamma - 1$)
would be larger than 1. 
Therefore, the links or degree of the largest hub 
grows faster than the size of the network \cite{Barabasi_2013_Network_Science}.
Hence, a cut-off point that will be reached and the largest 
hub will not grow any larger. 
These cut-offs possibly exist in 
STEs in GPP due to continental discontinuity of 
terrestrial GPP \cite{Zscheischler_2013_STE}. 
The networks with $1 < \gamma < 2$ are categorized
as anomalous regimes \cite{Barabasi_2013_Network_Science} 
in which large networks cannot exist.

\vspace{.5cm}
\subsection{\textbf{Attribution of Negative STEs in GPP to Climate Drivers}}

We selected the 100 largest STEs in GPP for attribution to 
\emph{tas} and \emph{pr} anomalies. 
We also analyzed the impact of antecedent climatic conditions
on negative STEs in GPP and found that as the lagged number of months
increases, the number of STEs attributed to a given climatic variable
decreases. 
However, for ease of representing drivers, we reported 
the average number of STEs driven from a lag of zero to 
three months. 
Table~\ref{t:attr_fluxann} shows the number of 
negative STEs in GPP using FluxANN driven by various climate drivers from ERA5,
and Table~\ref{t:attr_cesm2} shows attribution of STEs in GPP to
climate drivers from CESM2.
The dominant climate driver of negative STEs in GPP was the
\emph{hot} climatic condition for both FluxANN and CESM2, followed
by the \emph{cold} climatic condition. 
The dominance of \emph{hot} events on negative STEs in 
GPP was seen for all neighborhood structures.
However, the dominance of \emph{hot} events driving negative
STEs declined as the number of neighbors in neighborhood structures
increased in CESM2; for \emph{leld} the attribution to 
\emph{hot}:\emph{cold} was 50:50.
The number of STEs driven by \emph{dry} and \emph{wet} were
very rare.
These attribution results were in contrast to the findings 
from other studies \cite{Zscheischler_2014_IAV_GPP,
Zscheischler_2013_STE,Reichstein_2013_climate_ext,
Zscheischler_GRL_2014, Sharma_2022_CarbonExtremesLULCC,
Sharma_2022_NBPExtremes}, which requires further investigation.
Main difference between this attribution analysis and prior work
was identified that could have driven differences in our results.
Recent studies \cite{Sharma_2022_CarbonExtremesLULCC,Sharma_2022_CarbonExtremesLULCC}
have performed attribution analysis at every grid and 
a few studies \cite{Zscheischler_2013_STE,Zscheischler_GRL_2014} have 
compared the left and right tails of distribution of climate drivers 
based of a moving window of STEs in GPP relative to the median of 
the climate drivers during STEs in GPP. 
Here, we represented the median climatic condition for the whole manifold of
STEs in GPP at antecedent time steps and compare it with a median climatic condition
during a STE event in GPP.

\begin{table}[]
\centering
\caption{Attribution of negative carbon cycle STEs from FluxANN to monthly average surface temperature and precipitation from ERA5. 
The first and fourth quartile values of drivers were used to define cold, hot, dry, and wet climatic conditions. 
The attribution analysis is conducted for top 100 STEs. 
The values represent average response of STEs to climate drivers from lag of zero months to three months.}
\label{t:attr_fluxann}
\resizebox{0.7\columnwidth}{!}{%
\begin{tabular}{|l|r|r|r|r|}
\hline
\textbf{FluxANN} & \textit{cold} & \textit{hot} & \textit{dry} & \textit{wet} \\ \hline
\textit{seld}             & 3             & 13           & 4           & 1            \\ \hline
\textit{lesd}             & 4            & 18           & 2            & 1            \\ \hline
\textit{6-n}              & 1             & 10           & 0            & 0            \\ \hline
\textit{18-n}             & 2             & 11           & 0            & 0            \\ \hline
\textit{leld}             & 1             & 10           & 0            & 0            \\ \hline
\end{tabular}%
}
\end{table}

\begin{table}[]
    \centering
    \caption{Attribution of negative carbon cycle STEs from FluxANN to monthly average surface temperature and precipitation from ERA5. 
    The first and fourth quartile values of drivers were used to define cold, hot, dry, and wet climatic conditions. 
    The attribution analysis is conducted for top 100 STEs. 
    The values represent average response of STEs to climate drivers from lag of zero months to three months.}
    \label{t:attr_cesm2}
    \resizebox{.7\columnwidth}{!}{%
    \begin{tabular}{|l|r|r|r|r|}
    \hline
    \textbf{CESM2} & \textit{cold} & \textit{hot} & \textit{dry} & \textit{wet} \\ \hline
    \textit{seld}           & 0             & 38          & 0            & 1            \\ \hline
    \textit{lesd}           & 7            & 19           & 1            & 3           \\ \hline
    \textit{6-n}            & 5            & 9           & 1            & 0            \\ \hline
    \textit{18-n}           & 4            & 7           & 0            & 0            \\ \hline
    \textit{leld}           & 5            & 5           & 0            & 1           \\ \hline
    \end{tabular}%
    }
    \end{table}

Since the distribution of STEs in GPP follow a power law, 
the top 100 STEs represent very large extent and long duration events.
Our aim was to compute representative climatic condition for each driver 
during and prior to STEs.
One drawback of this approach is that summarizing the climatic 
conditions to one value per STE 
cannot represent the regional heterogeneity of climate--carbon cycle feedbacks.

\vspace{1cm}

\section{Conclusions}

The comparative analysis of carbon cycle STEs among observation (FluxANN)
and ESM (CESM2) indicated that:
\begin{enumerate}
    \item The magnitude of losses in carbon uptake during negative 
    STEs in GPP were larger in CESM2 than FluxANN. One of the main 
    reason of underestimating the magnitude of negative carbon cycle 
    extremes in FluxANN is due to low variability in carbon cycle fluxes.
    \item The largest magnitude of negative carbon cycle extremes
    were in the tropical regions, especially in the Amazon Basin. 
    \item The area affected and the magnitude of carbon losses during negative STEs in GPP
    were largely dependent and proportional to the neighborhood structure 
    of STEs.
    \item More than 75\% of cumulative carbon loss during negative STEs in GPP
    was represented by less than 100 STEs.
    \item The largest driver of negative STEs in GPP was \emph{hot} climatic 
    conditions.
    \item As the size of the neighborhood structure increased, the attribution
    of negative STEs in GPP to \emph{hot}:\emph{cold} reduced. Further analysis is 
    needed to understand these patterns.
\end{enumerate}

Further analysis is needed to investigate the reasons for fewer attribution 
of negative STEs in GPP to \emph{dry} climatic conditions, which has been found to 
be the dominant driver of negative carbon cycle extremes in other studies 
\cite{Reichstein_2013_climate_ext,Sharma_2022_NBPExtremes,
Sharma_2022_CarbonExtremesLULCC}.
This study aimed to inform the community that the choice of neighborhood structure 
to find the spatiotemporal manifold in carbon cycle extremes governs the characteristics of 
continuous, constrained continuous, and non-continuous extremes 
that vary largely in magnitude and the area affect by extremes 
as well as the attribution to climate drivers.

\section*{Acknowledgment}
We acknowledge the World Climate Research Programme, which, through its Working Group on Coupled Modelling,
coordinated and promoted CMIP6. We thank the climate modeling groups for producing and making available 
their model output, the Earth System Grid Federation (ESGF) for archiving the data and providing access, 
and the multiple funding agencies who support CMIP6 and ESGF. 
We thank DOE's RGMA program area, the Data Management program, and NERSC for making this 
coordinated CMIP6 analysis activity possible.
The authors from ORNL are supported by the U.S. Department of Energy, Office of Science, Office of Biological and Environmental Research. ORNL is managed by UT-Battelle, LLC, for the DOE under contract DE-AC05-00OR22725.

\section*{Open Research}

The details of the data and its availability is mentioned 
in the Section ~\ref{sec:Data}.

This  data analysis was performed in Python, and the analysis codes 
are available on GitHub at 
\href{https://github.com/sharma-bharat/Codes_SpatioTemporalExtremes.git}
{https://github.com/sharma-bharat/Codes\_SpatioTemporalExtremes.git}.

\end{document}